# Light beam dynamics in materials with radially-inhomogeneous thermal conductivity

Yaroslav V. Kartashov,[1,2,*] Victor A. Vysloukh,[3,1] and Lluis Torner[1]

[1]*ICFO-Institut de Ciencies Fotoniques, and Universitat Politecnica de Catalunya, 08860 Castelldefels (Barcelona), Spain*
[2]*Institute of Spectroscopy, Russian Academy of Sciences, Troitsk, Moscow Region, 142190, Russia*
[3]*Departamento de Fisica y Matematicas, Universidad de las Americas – Puebla, Santa Catarina Martir, 72820, Puebla, Mexico*

We study the properties of bright and vortex solitons in thermal media with nonuniform thermal conductivity and homogeneous refractive index, whereby the local modulation of the thermal conductivity strongly affects the entire refractive index distribution. While regions where the thermal conductivity is increased effectively expel light, self-trapping may occur in the regions with reduced thermal conductivity, even if such regions are located close to the material boundary. As a result, strongly asymmetric self-trapped beams may form inside a ring with reduced thermal conductivity and perform persistent rotary motion. Also, such rings are shown to support stable vortex solitons, which may feature strongly non-canonical shapes.

Thermal self-action of light beams is a topic of continuously renewed interest. It may lead to rich evolution dynamics even in apparently simple uniform media, because the geometry and transverse dimensions of the sample define not only the characteristic scales of the heat flow, but also the functional form of the optical response (aka Green function) of the material. The effect was demonstrated in the very early papers on thermal self-action of laser beams [1,2].

Thermal nonlinearity can support and stabilize a rich variety of nonlinear self-trapped waves, with the possibility to control their dynamical evolution by adjusting the boundary conditions. In particular, stable two-dimensional fundamental and vortex solitary waves [3-7] can be readily generated in focusing thermal media [6]. Because of the very nature of the thermal nonlinearity, the interactions, or nonlinear coupling, between such beams may exhibit an unlimited long range [8]. Interaction forces between the self-trapped beams may lead to the formation of multi-spot stable patterns, in the form of multipole solitons [9-13]. The boundary conditions affect the beam trajectory [14] and determine the shape of the entire induced refractive index landscape [15]. If the thermal nonlinearity is defocusing, then light is expelled toward the surface of the sample, where stationary surface waves may form with shapes that are dictated by the geometry of the material sample [16].

By and large, structuring the optical properties of materials enriches the families of nonlinear waves available, often in non-intuitive ways. For example, in layered structures fabricated using alternating focusing and defocusing layers of thermal media, stable bright solitons were shown to exist even when the averaged nonlinearity of the material is defocusing [17]. Also, at the interface between a layered thermal medium and a linear dielectric, formation of stable surface multipole light patterns becomes possible [18].

Heat flux manipulation in engineered thermal materials, thus with an inhomogeneous thermal conductivity, have recently reached a new level, whereby suitable structures have been show to shield, to concentrate and even to invert the heat flux [19]. As a consequence, several previously elusive effects, such as thermal cloaking, formation of thermal lattices, controllable heat concentration and its imaging became accessible and thus have been experimentally demonstrated [20,21]. Similarly, the technique of complex heat flux generation opens up the possibility to explore the self-action of light beams in such media.

In this Letter we address the formation of fundamental (bright) and vortex soliton light states in media with non-uniform thermal conductivity and uniform refractive index. Specifically, here we consider rectangular samples where the conductivity is locally increased or decreased within a ring surrounding the central region. We found that strongly asymmetric bright self-trapped states may exist inside the ring with reduced thermal conductivity. Also, such rings are shown to support stable vortex solitons, which become strongly asymmetric in the presence of a global heat flowing from the heated facet of the sample to the cold one.

In steady-state regime, the propagation of light beams along the $\xi$-axis of a focusing thermal medium with a transversally inhomogeneous thermal conductivity, can be described by the system of coupled equations for the dimensionless amplitude of the light field $q$ and the nonlinear contribution to the refractive index $n$:

$$i\frac{\partial q}{\partial \xi} = -\frac{1}{2}\left(\frac{\partial^2 q}{\partial \eta^2} + \frac{\partial^2 q}{\partial \zeta^2}\right) - nq,$$
$$\frac{\partial \kappa}{\partial \eta}\frac{\partial n}{\partial \eta} + \frac{\partial \kappa}{\partial \zeta}\frac{\partial n}{\partial \zeta} + \kappa\left(\frac{\partial^2 n}{\partial \eta^2} + \frac{\partial^2 n}{\partial \zeta^2}\right) = -|q|^2. \quad (1)$$

Here $\eta,\zeta$ are the normalized transverse coordinates, $\xi$ is the propagation distance scaled to the diffraction length, and the function $\kappa(\eta,\zeta)$ describes the distribution of the thermal conductivity inside the sample. The nonlinear contribution to the refractive index $n$ is proportional to the local variation of the temperature in a given spatial point. The normalizations of all quantities are identical as those introduced in Ref. [10]. The system of Eqs. (1) has to be completed with the boundary conditions at the edges of a square sample with size $L \times L$. We assume that the intensity of the light field vanishes on all sample boundaries $|q|_{\eta,\zeta=\pm L/2}=0$, while the refractive index distribution is dictated by the temperature at which the boundaries are maintained. In order to take into account the potential role of temperature gradients, we set $n_{\eta=-L/2}=n_{\text{left}}$, $n_{\eta=+L/2}=n_{\text{righ}}$, and assume that at the two other boundaries the refractive index varies linearly following the expression $n_{\zeta=\pm L/2}=(n_{\text{righ}}-n_{\text{left}})(\eta/L)+(n_{\text{righ}}+n_{\text{left}})/2$. One may always set $n_{\text{righ}}=0$ and consider the case $n_{\text{left}} \geq 0$ without losing generality. The thermal conductivity profile is described by the function $\kappa(\eta,\zeta)=1+\delta\kappa \exp\{-[(\eta^2+\zeta^2)^{1/2}-r]^4\}$, where $r$ is the radius of the narrow ring surrounding the axis of the sample, so that $\kappa(\eta,\zeta) \to 1+\delta\kappa$ within the narrow annular region, while outside this region (in the center of the sample and at its periphery) the conductivity takes on the background normalized value $\kappa=1$. Here we consider the amplitudes of modulation $\delta\kappa \in [-1,+1]$ and set $L=30$, which corresponds to a typical sample with a cross-section of $1.5 \times 1.5$ mm$^2$.

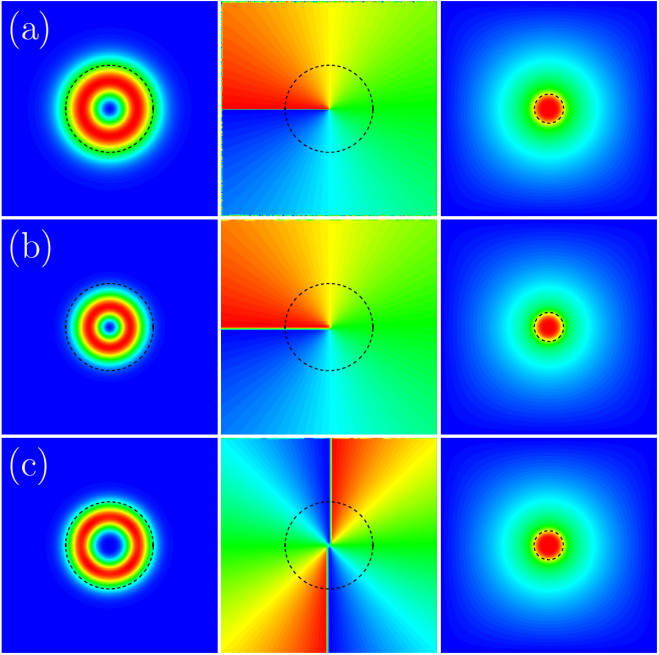

Fig. 1. Intensity (left), phase (center), and refractive index (right) distributions for vortex solitons in the inhomogeneous thermal medium at (a) $b=1.8$, $m=1$, (b) $b=5.7$, $m=1$, and (c) $b=5.7$, $m=2$. In all cases $\delta\kappa=-0.7$. The dashed line indicates the center of the ring with radius $r=2$, where the thermal conductivity is reduced. Notice the different scales in the right and the left panels.

Light propagation inside a thermal medium is accompanied by weak absorption, which acts as a heat source. Thermal conductivity results in the redistribution of such heat inside the entire sample. The final transverse steady-state temperature profile depends on the boundary conditions and it is much wider than the heating beam.

In materials with positive thermo-optic coefficient, the refractive index increases in regions with higher temperature, thus the self-induced waveguide that forms may support self-trapped light beams. Those can be written in the form $q=w(\eta,\zeta)\exp(ib\xi)$ in the case of fundamental beams and $q=[w_r(\eta,\zeta)+iw_i(\eta,\zeta)]\exp(ib\xi)$ in the case of vortex beams, where in both cases $b$ is the propagation constant. Examples of vortex solitons with various topological charges $m$, residing in the center of the sample with modulated thermal conductivity in the absence of temperature gradients ($n_{\text{left}}=0$), are shown in Fig. 1. In this paper we report mostly the results obtained for such solutions, because the properties of the fundamental solitons were found to be qualitatively similar. As it is visible in the plots, the refractive index profile is much wider than the vortex ring. For any modulation depth of the thermal conductivity $\delta\kappa$ the refractive index exhibits a nearly flat-top profile within the region occupied by the vortex soliton. The value of $\delta\kappa$ drastically affects the entire refractive index distribution: for negative $\delta\kappa$ (especially for $\delta\kappa \to -1$) the refractive index drops down abruptly outside the ring with reduced thermal conductivity, while for positive $\delta\kappa$ the refractive index smoothly decreases toward the edges. As expected, the self-trapped beams become narrower when the propagation constant increases; they may even concentrate completely in the inner region of the ring with the reduced $\kappa$ value, as readily visible by comparing Figs. 1(a) and 1(b). Note that this is in contrast to materials with modulated refractive index, where the supported solitons remain within the guiding ring when the propagation constant $b$ increases. At the same time, the variation of $\delta\kappa$ strongly affects the soliton shapes. At $\delta\kappa>0$ light is expelled into the regions located inside or outside the ring, depending on the value of the propagation constant, upon an increase of $\delta\kappa$. This indicates that the variation of $\kappa$ creates a pseudo-potential, akin to a nonlinear lattice, which is expulsive for $\delta\kappa>0$ and which enhances localization for $\delta\kappa<0$.

The properties of the solitons centered on the axis of the sample are described in Fig. 2. The soliton energy flow $U=\iint |q|^2 \, d\eta d\zeta$ monotonically decreases, while its integral width $W=2U^{-1}\iint (\eta^2+\zeta^2)^{1/2}|q|^2 \, d\eta d\zeta$ grows when the propagation constant $b$ decreases, as it is illustrated for the vortex soliton solution with topological charge $m=1$ [Fig. 2(a)]. For a fixed $b$, solitons are more localized and carry a smaller energy flow in a material with smaller $\delta\kappa$ [Fig. 2(b)]. The energy flow saturates at a constant level when $\delta\kappa \to +\infty$ because in this limit the beam is completely expelled from the ring with increased thermal conductivity. In contrast, the power carried by the solutions should vanish for any value of $b$ when $\delta\kappa \to -1$, which implies that the non-conducting ring embedded into the center of the sample prevents the formation of vortex solutions.

Vortex solitons in materials with the non-uniform thermal conductivity have complex stability domains. In Fig. 2(a) the unstable branches are shown in red, while stable branches are shown in black. For $\delta\kappa=-0.7$ the approximate borders of the stability domains (defined using direct propagation of slightly perturbed solitons) are given by $1.5<b<5.4$. At $\delta\kappa=-0.8$ the stability window is found to occur for

$1.2 < b < 2.6$, while at $\delta\kappa \to -1$ the entire soliton family becomes unstable. In contrast, the stability domain expands when $\delta\kappa \to 0$ and at $\delta\kappa = 0$ vortex solitons with $m=1$ become completely stable. Further growth of thermal conductivity was found to destabilize the vortex soliton solutions again. Although stability may be found even for $\delta\kappa = +1$, it was encountered only for small $b$ values. It should be stressed that fundamental solitons are always stable as long as the thermal conductivity is reduced within the central ring, but drift instabilities become possible for $\delta\kappa > 0$.

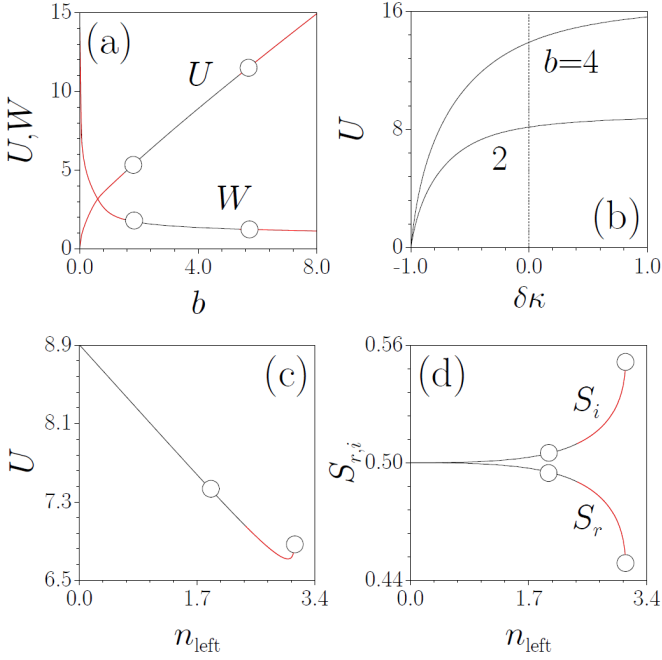

Fig. 2. (a) Energy flow $U$ and integral width $W$ of the vortex soliton solutions versus the propagation constant $b$ at $\delta\kappa = -0.7$. Circles correspond to the solutions shown in Figs. 1(a) and 1(b). (b) Energy flow of the vortex soliton solutions versus $\delta\kappa$ for different $b$ values. The dashed line corresponds to a uniform medium. Energy flow (c) and energy fractions concentrated in the real and imaginary parts of the field (d) versus $n_{\text{left}}$ at $b=4$, $\delta\kappa = -0.7$. Circles correspond to the solutions depicted in Fig. 3. In panels (a),(c), and (d) stable soliton branches are shown in black, while unstable branches are shown in red. In all cases, $m=1$.

The presence of an external temperature gradient results in notable asymmetries of the soliton profiles. Remarkably, the ring with the reduced thermal conductivity may prevent the drift of the beam toward the boundary with higher temperature and is capable to hold vortices around the center of the sample, provided that temperature gradient does not exceed a critical value (Fig. 3). When the temperature gradient increases, the vortex solutions acquire strongly non-canonical shapes. In contrast to intuitive expectations, the amplitude of the solutions is locally increased not in the direction of the temperature gradient, but in the perpendicular direction [Fig. 3(b)]. The phase dislocation shifts toward the boundary that exhibits a higher temperature. The refractive index distribution is also strongly asymmetric [compare Fig. 3(c) showing the refractive index profile in the presence of a vortex beam, with Fig. 3(d) showing the refractive index distribution that occurs in the absence of the heating beam].

The presence of the temperature gradient results also in the appearance of an energy flow threshold for soliton formation [Fig. 2(c)]. As visible in the plots, soliton solutions were not found above a maximal $n_{\text{left}}$ value (note that we consider the case $n_{\text{left}} \geq 0$), which was found to depend on both, $\delta\kappa$ and $b$. Vortex soliton solutions were found to be stable when $0 < n_{\text{left}} < n_{\text{cr}}$, as long as the refractive index gradient does not exceed a critical value $n_{\text{cr}}$. The distortion of the shape of the non-canonical vortices is obvious from comparison of the power fractions $S_{r,i} = U^{-1}\iint w_{r,i}^2 d\eta d\zeta$ concentrated in the imaginary and real parts of the field, since in the uniform medium $S_r = S_i$, depicted in Fig. 2(d).

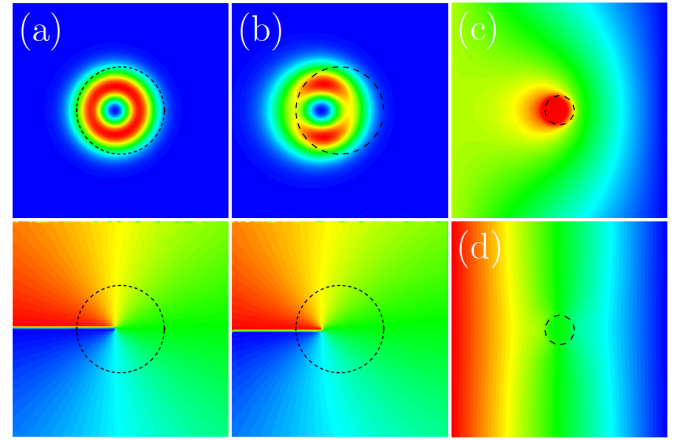

Fig. 3. Intensity and phase distributions for vortex solitons with $m=1$, $b=4$, $\delta\kappa = -0.7$ at $n_{\text{left}} = 2.0$ (a) and $n_{\text{left}} = 3.09$ (b). Panel (c) shows the refractive index distribution induced by the vortex soliton depicted in panel (b). Panel (d) shows the refractive index distribution that occurs in the absence of any heating beam, for $\delta\kappa = -0.9$, $n_{\text{left}} = 3.0$. Notice the different scales in panels (a),(b) and panels (c),(d).

One of the central results of this Letter is that soliton solutions in media with inhomogeneous thermal conductivity may form not only in the center of the sample (which is the only point of stable equilibrium in the uniform thermal medium), but also within the region with reduced $\kappa$ value, even if it is located close to the sample boundary. Stable fundamental solitons exist not only on localized thermal conductivity defects, but also inside the ring with reduced conductivity. If the value of $\delta\kappa$ is sufficiently small, for example $\delta\kappa = -0.9$, such solitons reside inside the ring and exist above an energy flow threshold and above a cutoff of the propagation constant [Fig. 4(a)]. For $\delta\kappa > -0.7$ the cutoff disappears and upon decrease of the propagation constant the soliton leaves the ring and gradually falls on the center of the sample when $b \to 0$. This is accompanied by a considerable broadening of the soliton profile.

The refractive index distribution for solitons residing in the ring was found to be strongly asymmetric [Fig. 4(b)], with a pronounced maximum in the region with reduced thermal conductivity. Notice the rapid decrease of the refractive index at $|(\eta^2+\zeta^2)^{1/2}| > r$ (outside the ring with reduced thermal conductivity), that is also visible in the full refractive index distributions shown in Fig. 5. It is worth noting that soliton solutions residing in the ring satisfy the Vakhitov-Kolokolov stability criterion, i.e. they are stable in the $b$ interval where $dU/db > 0$ and are unstable otherwise.

In the presence of input phase tilts in the form $\exp(i\alpha_\eta \eta + i\alpha_\zeta \zeta)$, which introduce a linear beam velocity tangential to the ring, fundamental single-spot solutions perform persistent stable rotary motion, without any signatures of radiation or decay over hundreds of rotation periods (Fig. 5). Such robustness is a remarkable result, taking into account that in uniform thermal media boundaries always affect dramatically the trajectory of motion of tilted beam and that one may reasonably expect the presence of a so-called effective radiative friction associated with the square geometry of the sample (see Ref. [11]). Notice that the trajectory of motion of the integral soliton center, shown in Fig. 5, is not strictly circular but rather polygonal instead. The effect is more visible from trajectories obtained for large phase tilts. It should be also stressed that persistent rotation is possible not only for fundamental solitons, but also for stable complexes of out-of-phase beams trapped inside the ring with reduced $\kappa$ value.

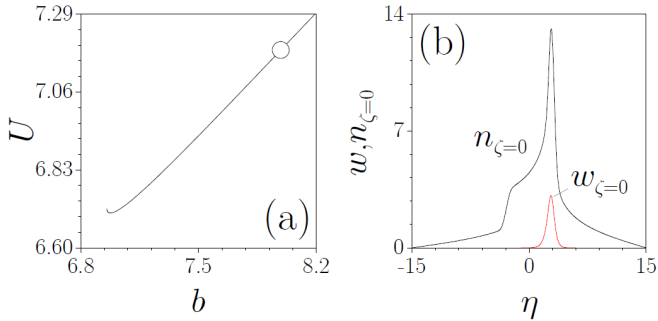

Fig. 4. (a) Energy flow $U$ versus $b$ for soliton trapped inside the ring with reduced thermal conductivity at $r = 3$, $\delta\kappa = -0.9$. (b) The cross-sections of the soliton and the refractive index at $\zeta = 0$ corresponding to the circle in (a).

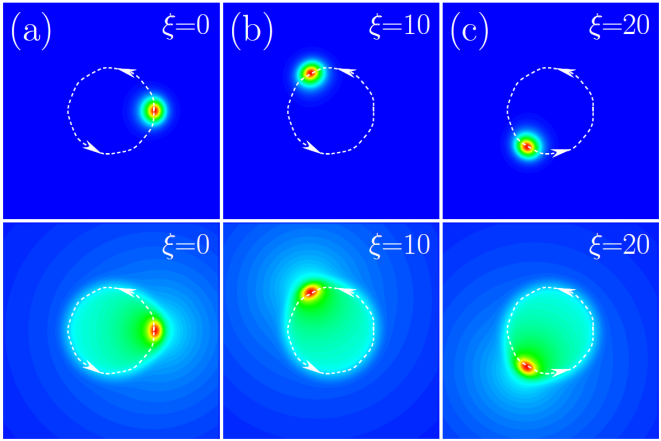

Fig. 5. Intensity (top row) and refractive index (bottom row) distributions at different distances showing stable rotation of the fundamental soliton with $b = 8$, $\delta\kappa = -0.9$ and initial phase tilt $\alpha_\eta = 0.0$, $\alpha_\zeta = 0.6$ inside the ring with reduced thermal conductivity and radius $r = 3$. The dashed line shows the trajectory of the soliton center, while the arrows show the direction of rotation and are simply to help the eye.

It is worth emphasizing the differences between the thermally trapped bright single-spot and vortex beams studied here and standard solitons that form in unbounded media with non-local nonlinearity (with, e.g., a Gaussian response function). In the latter case, the soliton parameters are spatially invariant and depend mainly on the characteristic scale of the response function, which is smooth over the intensity profile that forms the refractive index distribution. However, in our case the sample cross-section and the spatial distribution of the thermo conductivity dictate not only the shape and related features of the trapped beam, but also the position of its stable equilibrium.

In summary, we have shown that a spatial modulation of the thermal conductivity of suitable nonlinear media can drastically affect the light-induced refractive index profiles mediated by thermal nonlinearities and, as a consequence, can importantly impact the properties of propagating bright single-spot and vortex self-trapped soliton light states. In particular, such geometries afford trapping off-center light beams and the generation of rich light beam dynamics, such as persistent rotation of self-trapped light states over many diffraction lengths.